\documentclass[journal]{IEEEtran}

\usepackage{booktabs}

\usepackage[hidelinks]{hyperref}
\usepackage{graphicx}
\graphicspath{{images/}}
\usepackage{multicol}
\usepackage{lipsum}
\usepackage{mathtools}
\usepackage{amsthm}
\usepackage{amsmath,amssymb,amsfonts}
\usepackage{stfloats}
\usepackage{amsfonts}
\usepackage{acronym}
\usepackage{cite}
\usepackage{multirow}
\usepackage{bm}
\usepackage{amsmath,amssymb}
\usepackage{graphicx}
\usepackage[table,xcdraw]{xcolor}
\usepackage[english]{babel}
\ifCLASSOPTIONcompsoc
 \usepackage[caption=false,font=normalsize,labelfont=sf,textfont=sf]{subfig}
\else
 \usepackage[caption=false,font=footnotesize]{subfig}
\fi
\usepackage{caption}
\usepackage{subcaption}
\usepackage[geometry]{ifsym}
\usepackage{array}
\usepackage[ruled,vlined]{algorithm2e}
\usepackage[nowarn,acronyms,nonumberlist,nopostdot,nomain,nogroupskip]{glossaries}
\usepackage{fancyhdr}

\newcommand\submittedtext{%
  \footnotesize This work has been submitted to the IEEE Communications Magazine for possible publication. Copyright may be transferred without notice, after which this version may no longer be accessible.}

\fancypagestyle{submittednotice}{
  \fancyhf{} 
  \fancyhead[C]{\fbox{\parbox{\dimexpr0.75\textwidth-\fboxsep-\fboxrule\relax}{\submittedtext}}} 
}

\newacronym{3gpp}{3GPP}{3rd Generation Partnership Project}
\newacronym{5g}{5G NR}{fifth generation new radio}
\newacronym{6g}{6G}{sixth generation}
\newacronym{5gc}{5GC}{5G core}
\newacronym{ai}{AI}{artificial intelligence}
\newacronym{ar}{AR}{augmented reality}
\newacronym{api}{API}{application programming interface}
\newacronym{achem}{ACHEM}{AERPAW Channel Emulator}
\newacronym{aerpaw}{AERPAW}{Aerial Experimentation and Research Platform for Advanced Wireless}
\newacronym{av}{AV}{autonomous vehicle}
\newacronym{avn}{AVN}{autonomous vehicle network}
\newacronym{bs}{BS}{base station}
\newacronym{chem}{CHEM}{channel emulator}
\newacronym{dac}{DAC}{digital-to-analog converter}
\newacronym{darpa}{DARPA}{Defense Advanced Research Projects Agency}
\newacronym{dl}{DL}{downlink}
\newacronym{dt}{DT}{digital twin}
\newacronym{dtn}{DTN}{digital twin network}
\newacronym{enb}{eNB}{E-UTRAN NodeB}
\newacronym{epc}{EPC}{Evolved Packet Core}
\newacronym{fifo}{FIFO}{first-in-first-out}
\newacronym{fpga}{FPGA}{field programmable gate array}
\newacronym{gnb}{gNB}{Next-Generation Node B}
\newacronym{hitl}{HITL}{hardware-in-the-loop}
\newacronym{iot}{IoT}{Internet-of-Things}
\newacronym{ioe}{IoE}{Internet-of-Everything}
\newacronym{itu-t}{ITU-T}{ITU Telecommunication Standardization Sector}
\newacronym{lte}{LTE}{Long-Term Evolution}
\newacronym{ng}{NextG}{next-generation}
\newacronym{ntn}{NTN}{non-terrestrial network}
\newacronym{mmtc}{mMTC}{massive machine-type communications}
\newacronym{otw}{OTW}{over-the-wire}
\newacronym{oran}{O-RAN}{open radio access network}
\newacronym{pci}{PCI}{physical cell identity}
\newacronym{prb}{PRB}{physical resource block}
\newacronym{pt}{PT}{physical twin}
\newacronym{p2v}{P2V}{Physical-to-Virtual}
\newacronym{rf}{RF}{Radio Frequency}
\newacronym{rw}{RW}{real-world}
\newacronym{rsrp}{RSRP}{reference signal received power}
\newacronym{rssi}{RSSI}{received signal strength index}
\newacronym{sdr}{SDR}{software-defined radio}
\newacronym{snr}{SNR}{signal-to-noise ratio}
\newacronym{sitl}{SITL}{software-in-the-loop}
\newacronym{vr}{VR}{virtual reality}
\newacronym{uav}{UAV}{uncrewed aerial vehicle}
\newacronym{ue}{UE}{user equipment}
\newacronym{ul}{UL}{Uplink}
\newacronym{ugv}{UGV}{uncrewed ground vehicle}
\newacronym{usrp}{USRP}{universal software radio peripheral}
\newacronym{v2p}{V2P}{Virtual-to-Physical}
\newacronym{xr}{XR}{extended reality} 
\newacronym{qos}{QoS}{quality-of-service}
\newacronym{oac}{OAC}{over-the-air computation}
\newacronym{fl}{FL}{federated learning}

\ifCLASSINFOpdf
\else
\fi

\begin{document}

\title{Digital Twins and Testbeds for Supporting AI Research with Autonomous Vehicle Networks}


\author{An\i l G\"urses,
Gautham~Reddy,
Saad~Masrur,
\"Ozg\"ur \"Ozdemir,
\.{I}smail~G\"uven\c{c},
Mihail~L.~Sichitiu,
Alphan~\c{S}ahin,
Ahmed~Alkhateeb,
Magreth Mushi,
Rudra~Dutta
\thanks{This work is funded in part by NSF award CNS-1939334, and the datasets used in numerical results are available publicly at \cite{aerpaw_website}.}}

\markboth{IEEE Communications Magazine}%
{ \MakeLowercase{\textit{et al.}}: Digital Twins and Testbeds for Supporting AI Research with Autonomous Vehicle Networks}

\maketitle
\thispagestyle{submittednotice}

\begin{abstract}
    Digital twins (DTs), which are virtual environments that simulate, predict, and optimize the performance of their physical counterparts, hold great promise in revolutionizing next-generation wireless networks.
    While DTs have been extensively studied for wireless networks, their use in conjunction with autonomous vehicles featuring programmable mobility remains relatively under-explored.
    In this paper, we study DTs used as a development environment to design, deploy, and test artificial intelligence (AI) techniques that utilize real-world (RW) observations, e.g. radio key performance indicators, for vehicle trajectory and network optimization decisions in autonomous vehicle networks (AVN).
    We first compare and contrast the use of simulation, digital twin (software in the loop (SITL)), sandbox (hardware-in-the-loop (HITL)), and physical testbed (PT) environments for their suitability in developing and testing AI algorithms for AVNs.
    We then review various representative use cases of DTs for AVN scenarios. Finally, we provide an example from the NSF AERPAW platform where a DT is used to develop and test AI-aided solutions for autonomous unmanned aerial vehicles for localizing a signal source based solely on link quality measurements.
    Our results in the physical testbed show that SITL DTs, when supplemented with data from RW measurements and simulations, can serve as an ideal environment for developing and testing innovative AI solutions for AVNs.

\end{abstract}


\IEEEpeerreviewmaketitle

\section{Introduction}

The pervasiveness of digitization has propelled the emergence of a concept known as \gls*{dt} in a wide range of application areas in the last decade.
The emerging importance of \glspl*{dt} stems from their transformative impact on data-driven decision-making processes within the interconnected systems, enabling the integration of multiple data sources at single point~\cite{dtn_offloading}.
\Glspl*{dt} have gained significant attention recently for their potential in designing, evaluating, and testing wireless networks, particularly in addressing the complexities inherent in the development and validation of these systems in \gls*{rw} environments~\cite{dtn_6g_vision}.
Standardization bodies, such as \gls*{3gpp} and \gls*{itu-t}, have also recognized the importance of \glspl*{dt}~\cite{itu-tDTN}. 

As \gls*{3gpp} nears completion of \gls*{5g} standard, researchers are already eyeing the development of the next generation of wireless networks.
The \gls*{6g} is expected to support an even broader range of applications, including more advanced \glspl*{av}, specifically \glspl*{uav} and \glspl*{ugv}, along with \gls*{ai} integrations. 
\Glspl*{dtn} are poised to play a critical role in enabling this new wave of applications, which has the potential to revolutionize various industries.
As \glspl*{dt} progressively advance, they possess the potential to transform the development and evaluation process of vertical use cases of wireless networks by incorporating \gls*{ai} techniques.
For example, \gls*{dt}-enabled development environment can train, validate, and test \gls*{ai} algorithms for \glspl*{uav} in \glspl*{avn} without concerns related to airspace safety or wireless spectrum access rights. 
With the growing prevalence of \gls*{ai} across various applications, the ability to intelligently adapt \gls*{av} trajectories and wireless network configurations in response to changing environmental and network conditions is becoming increasingly crucial.
However, 
the current stage of \gls*{dt} technology in facilitating research and development in these areas remains in its infancy. 

\begin{figure}
    \centering
    \includegraphics[width=0.95\linewidth]{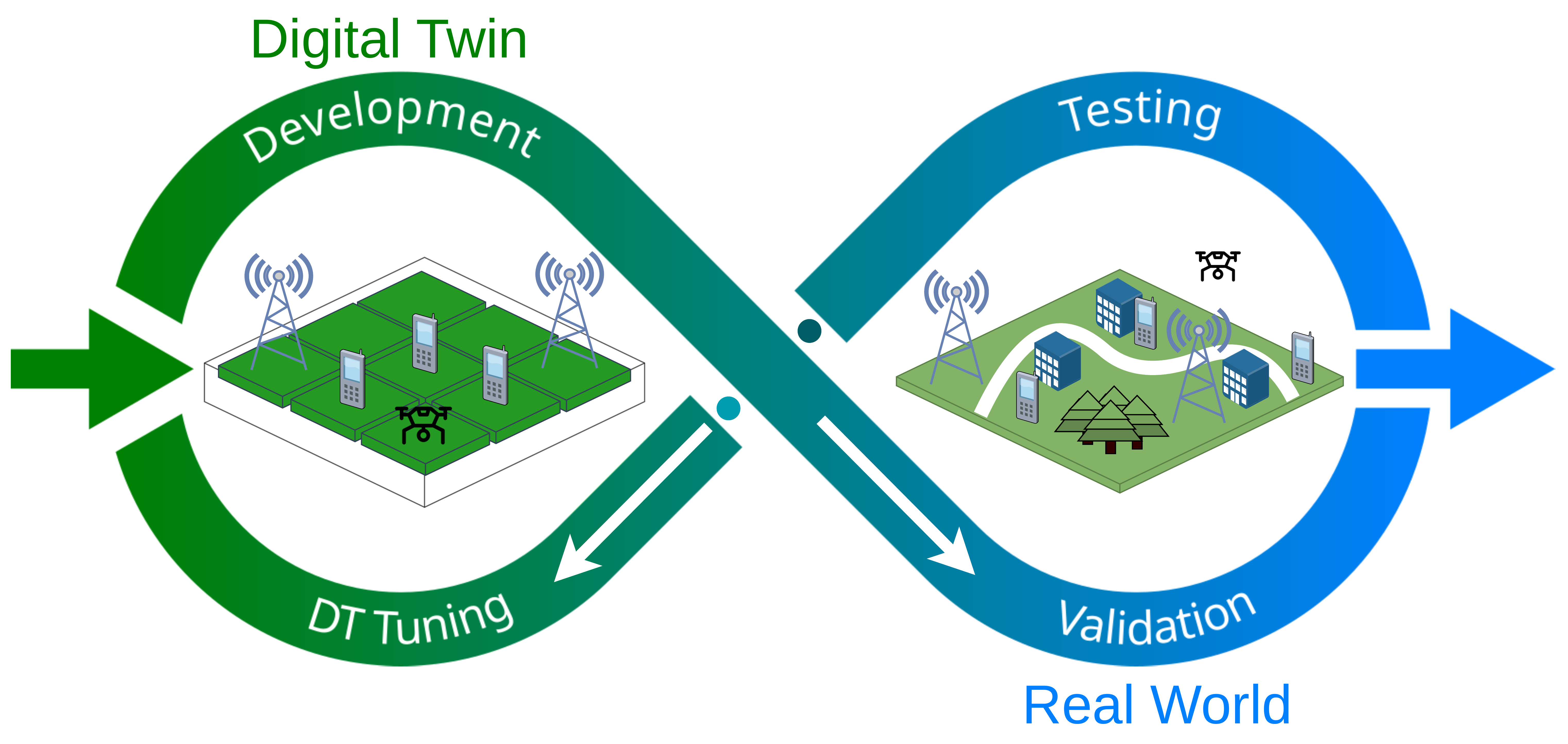} 
    \caption{DT assisted AVN technology development cycle from idea to \gls*{rw} deployment, including continuous refinement of DT through validation and tuning.}
    \label{fig:DT-Overview}\vspace{-6mm}
\end{figure}

A critical requirement of \glspl*{dt} is to not only have a strong \gls*{p2v} connection where a physical environment is accurately characterized in the virtual environment but also a strong \gls*{v2p} connection as depicted in Fig.~\ref{fig:DT-Overview}, where development of different network components in the virtual environment can be seamlessly transferred into the physical world~\cite{jones2020characterising}.
This interplay between the two environments requires a high-fidelity representation of the physical environment in the virtual environment. 
However, this requirement poses a significant challenge due to the complexity introduced by \glspl*{avn}~\cite{dtn_challenges}.
Research and development of a realistic \glspl*{dt} for \glspl*{avn} with \gls*{p2v} and \gls*{v2p} features require addressing multiple challenges. 
First, a \gls*{dt} should support the capability of developing and testing \emph{\gls*{rw}} \gls*{av}, specifically \gls*{uav} and \gls{ugv}, and radio software in a virtualized environment, so that intricate software interactions and constraints involving \glspl*{av} and radio nodes are captured. 
Second, a \gls*{dt} should be accompanied by a \gls*{rw} \gls*{pt} testbed in order to be able to validate \gls*{p2v} and \gls*{v2p} connections.
Having a tightly coupled \gls*{dt} and \gls*{pt} is also crucial for seamlessly moving experiments back and forward between the two environments for continuous improvement of a new idea.
Finally, it is crucial to realistically model the 3D \gls*{rw} propagation conditions in the \gls*{dt} at the frequencies of interest, especially when training new \gls*{ai} techniques in the \gls*{dt} environment prior to deployment in \gls*{rw}. 

The main goal of our work is to address the aforementioned challenges associated with developing \glspl*{avn}.
In this paper, we first present a comparative overview of development environments for \glspl*{avn}, including the simulation, \gls*{dt}, sandbox, and \gls*{pt}.
Next, we present representative \gls*{avn} scenarios where \gls*{ai}-based solutions are critical for \gls*{av} trajectory and wireless network optimization, and elaborate how \glspl*{dt} can be used to develop such solutions.
We then delve into the details of \gls*{aerpaw} platform~\cite{aerpaw_website} and their corresponding \gls*{dt}-enabled wireless research experimentation environment.
Finally, we present a case study on \gls*{ai}-aided signal source localization, where the solution approach is first developed in \gls*{aerpaw}'s \gls*{dt}, calibrated using \gls*{rw} data, and seamlessly moved for testing in the \gls*{pt} testbed by moving software containers to the testbed. 
Our representative results in the \gls*{dt} and \gls*{pt} environments show that DTs can significantly reduce costs and accelerate the \gls*{ai} development cycle for addressing \gls*{avn} challenges compared to development and testing exclusively in a testbed environment.  

\begin{figure*}[t]
    \centering
    \includegraphics[width=0.95\linewidth]{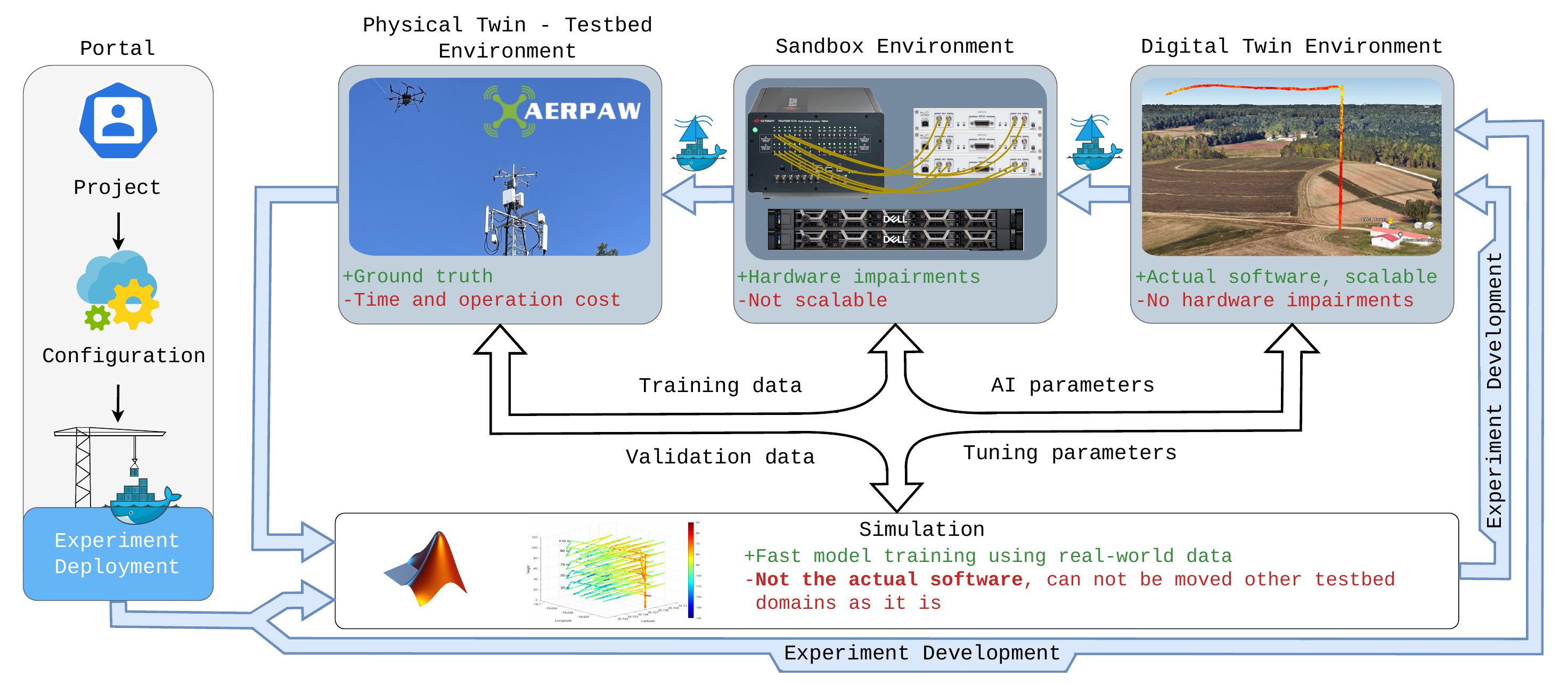}
    \caption{Interactions between \gls*{pt}, \gls*{dt}, sandbox, and simulation for experiment development and testing.}
    \label{fig:Testbed}\vspace{-4mm}
\end{figure*}

\begin{figure*}[t]
    \centering
    \includegraphics[width=0.95\linewidth]{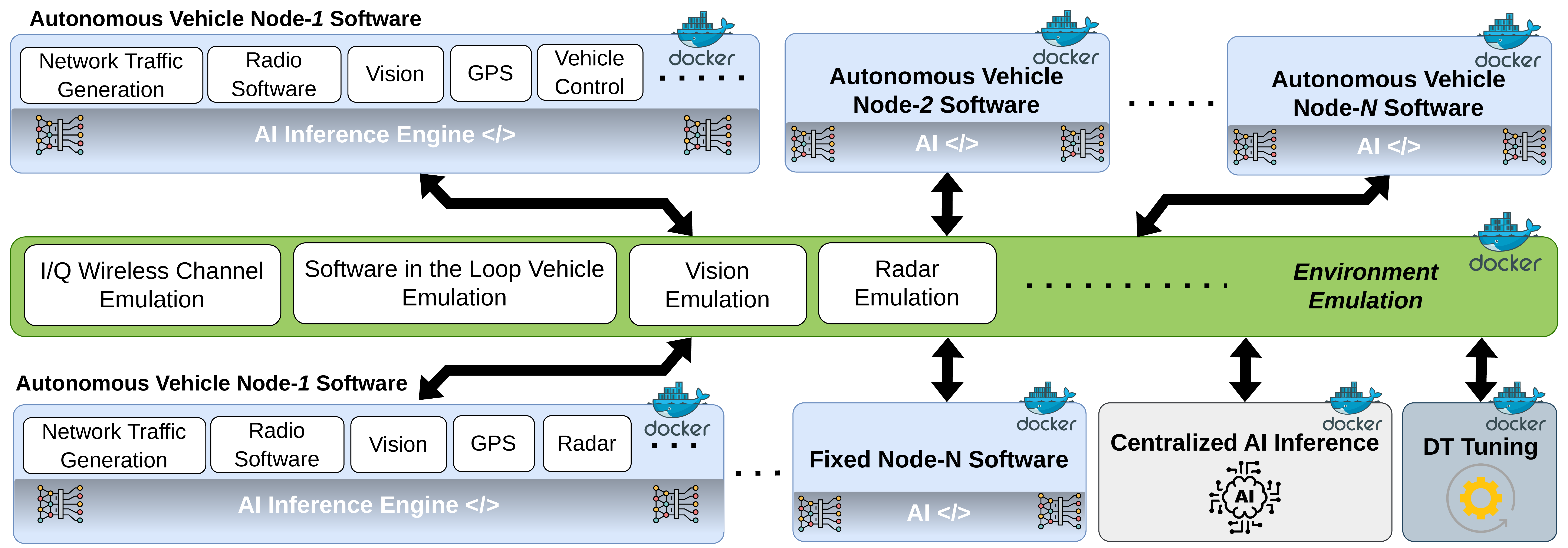}
    \caption{Interaction of various vehicle and fixed-node software through the environment emulator in the AVN-DT. AI inference engines at each vehicle and fixed node, as well as a centralized entity, are also illustrated.}
    \label{fig:DT_AI}\vspace{-4mm}
\end{figure*}

\section{Development Environments for Autonomous Vehicle Networks}

Although there are no established standards defined for \gls*{dt}-enabled networks yet, various potential applications have been envisioned. 
In the context of \glspl*{avn}, we envision three major use cases: 1) development and testing, 2) model training and validation for \gls*{ai} inference engines, and 3) real-time decision offloading.
This paper primarily focuses on the first two use cases of \glspl*{dt}. 
To facilitate these use cases, \glspl*{dt} must be able to engage with the physical realm, including the sandbox, \gls*{pt}, and simulations as shown in Fig.~\ref{fig:Testbed}.

\subsection{Development Using Simulations}

Simulations serve as an essential foundation for developing \gls*{ai} solutions for \glspl*{avn} by providing a controlled simplified environment for testing and validating the system.
Simulations utilize simplified representations of the system components to isolate and analyze key dynamics, primarily focusing on the specific component being developed.
With the flow envisioned by \glspl*{dt}, simulations can be made accurate and realistic, as the models are refined continuously based on the \gls*{rw} data collected from the system components in \gls*{pt}.
By enabling the modeling and analysis of key aspects of the system without having to build physical prototypes, simulations save time and resources while ensuring that only optimal solutions are passed into the implementation phase. 
Due to the abstractions used in the simulations, the development done for the simulations is not directly deployable to \gls*{rw}. 
For this specific reason, simulations are primarily used to anticipate system behavior while excluding some impairments that arise from implementation challenges.
As a result, while simulations are invaluable for theoretical understanding and analysis of \gls*{ai} aided \glspl*{avn}, they are only one part of a comprehensive development process that includes iterative refinement. 


\subsection{Development Using Digital Twins (SITL)}

The advancements in software development and computer hardware have led to the realization of complex virtual environments that can accurately mimic the behavior of physical systems, making it possible to emulate complex dynamics in digital space.
In this paper, we envision \glspl*{dt} that use environment emulators, as shown in Fig.~\ref{fig:DT_AI}, where actual \gls*{av}, radio, and other (e.g. vision) software interact with each other through the environment emulator, therefore capturing the complex protocol, radio, and vehicle control interactions in the \gls*{avn}. 
Although there are two types of emulators, \gls*{hitl} and \gls*{sitl} emulators, a complete \gls*{dt} can only be realized with \gls*{sitl}, with every component being digitally presented in a virtual environment, which brings significant advantages in experiment scalability, cost, and simultaneous use of the \glspl*{dt} by multiple users in a containerized platform.


The primary challenge with \gls*{sitl} emulators is the emulation's accuracy at various levels, such as wireless channel and vehicle mobility, which is impacted by the complexity of the environment and the system.
To tackle this challenge, we emphasize the significance of incorporating a \gls*{dt} tuning engine for calibration of the various models (e.g. channel and vision) based on \gls*{rw} data acquired from the \gls*{pt}.
The role of the \gls*{dt} tuning engine encompasses the analysis of \gls*{rw} data, the extraction of certain statistics, and the refining of the model deployed to the \gls*{sitl} emulator. 
Once the AI inference engines at individual nodes, as well as the centralized AI inference engine that may be interacting with all fixed and AV nodes, are fully developed and tested, the software containers can be moved seamlessly to the sandbox and \gls*{pt} environments for further testing and refinement. 
As a result, \glspl*{dt} can significantly reduce integration efforts for developing and testing \gls*{ai} solutions for \glspl*{avn} compared to development and testing exclusively in a simulation environment.

We want to note that emulators and simulators are often mistakenly thought to be the same.
However, they differ significantly in their approaches to system modeling: emulators use the genuine software of a system, whereas simulators create simplified models based on mathematical representations.
In the realm of \glspl*{dt}, employing multiple emulators operating on the \gls*{dt} (as depicted in Fig.~\ref{fig:DT_AI}) grants the ability to inspect protocol behavior at various levels, providing a deeper understanding of intricate protocol interactions.


\subsection{Development Using a Sandbox (HITL)}

As described in Fig.~\ref{fig:Testbed}, a sandbox development environment offers another alternative through the employment of actual hardware paired with either \gls*{hitl} channel emulators or isolated RF environments.
In \gls*{hitl} emulators, signals are transmitted via cable to the channel emulator, which emulates the wireless channel propagation, before being received by the receivers.
An example of this can be seen with \gls*{aerpaw}'s utilization of the Keysight Propsim channel emulator, as shown in Fig.~\ref{fig:Testbed}.
In contrast, isolated RF environments involve transmitting and receiving signals within a contained space, with path loss controlled by an attenuator to simulate varying wireless channel conditions.
Although isolated RF environments provide valuable insights into hardware impairments like those from radio front ends in SDRs, their scalability is limited.

\subsection{Development Using Physical Twins (\gls*{rw} Testbed)}

As a final development environment, \glspl*{pt}, which are \gls*{rw} environments equipped with actual hardware components in a testbed, can be used for the development of \gls*{ai} solutions for \glspl*{avn}.
At first blush, testbeds may seem to be the most appealing development environment due to their realistic operational characteristics that closely resemble \gls*{rw} deployments. 
However, testbeds are not scalable, flexible, and cost/time-effective, especially for \gls*{avn} experiments that require AV pilots, long testing time, and experimental spectrum licenses that are more restrictive for aerial vehicles, among other factors.
Specifically in the context of \gls*{ai} development, testbeds are not ideal for the development of AI inference engines, as they require a large number of experiments to be conducted to train and validate the AI models.
Consequently, testbeds are best suited for final system validation and for collecting \gls*{rw} data for \gls*{ai}-aided \glspl*{avn}.

\begin{figure*}[t]
    \centering
    \includegraphics[width=\linewidth]{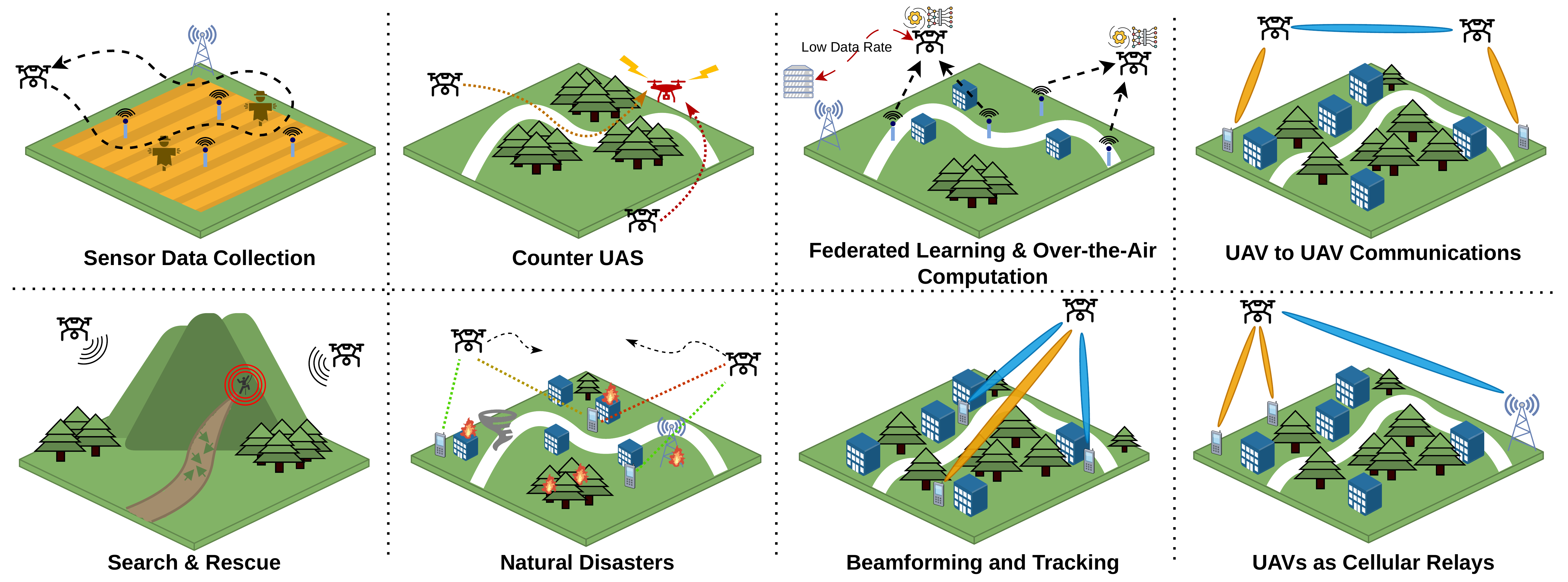}
    \caption{AVN application examples where AI techniques can be developed in a DT environment for \emph{jointly} optimizing AV trajectory control, swarm formation, and wireless network parameters before they get deployed in the RW.}
    \label{fig:application}\vspace{-4mm}
\end{figure*}

\section{Example Uses of AI with Digital Twins}

\glspl*{dtn} can enhance the efficiency and performance of the AI-aided \glspl*{avn} by leveraging the available virtual ground for training and validation, which allows for \emph{jointly} adapting the trajectories of the \glspl*{av} and the parameter configurations of the wireless network.
In this section, we provide several related use case examples as shown in Fig.~\ref{fig:application} and detail the interactions between the \gls*{ai} engines and the \glspl*{dtn}.

\subsection{Data Collection with UAVs}
\label{subsec:dataCollectionWithUAVs}
The use of \glspl*{uav} in areas with weak or non-existent cellular coverage, such as rural areas, has gained significant importance.
In agriculture systems, for instance, sensors spread across farms are used to improve crop yields and save costs in rural areas.
However, due to the large number of sensors and the size of the farms, it is not feasible to collect the data over-the-air from a single location.
To address this challenge, \glspl*{uav} can be deployed to fly along an optimized trajectory, efficiently collecting all the data from sensors in a timely manner~\cite{minimize_data_collection}.
Similar applications include using \glspl*{uav} to collect data from roadside units on roads, or from military sensors on a battlefield.
Although these applications can be optimized through various techniques, integrating \glspl*{dt} can further enhance the performance of the deployed \gls*{ai}-aided algorithms by enabling offline training to handle edge-case scenarios.
For instance, within the \gls*{dt}, a reinforcement learning model can be trained to optimize the \gls*{uav}'s flight trajectory for data collection, considering the wireless channel conditions and the UAV battery life.

\subsection{Disaster Recovery, and Search and Rescue with UAVs}\label{Sec:search_rescue}

Natural disasters such as earthquakes, wildfires, and hurricanes can cause significant damage to the infrastructure, which leads communication systems to be either overloaded or nonfunctional in affected areas.
In such scenarios, \glspl*{uav} can act as base stations (BSs) or relays to provide wireless connectivity and aid search and rescue teams.
Recovering the cellular coverage and locating those who are missing in these areas can be achieved by using \glspl*{uav} as moving BSs.
\gls*{ai} algorithms can optimize \glspl*{uav} placement and trajectory considering parameters such as signal quality and wireless coverage~\cite{uav_assisted_networks}.
With the integration of \gls*{dt} in the development cycle, the efficiency of \gls*{ai} algorithms can be improved, to better adapt \glspl*{uav} and network parameters to the environment, propagation, and other dynamics.  


\subsection{Counter UAS Applications}

One or more \glspl*{uav} equipped with \gls*{ai}-aided decision mechanisms can also be used to detect, localize, and track other signal sources, such as jammers in a dynamic spectrum-sharing environment, enemy units in a battlefield, or unauthorized \glspl*{uav} in an airspace. 
For example, a swarm of \glspl*{av} localizing and tracking unauthorized \glspl*{uav} is presented in~\cite{koohifar2018autonomous}. 
Once one of the tracking \glspl*{uav} approaches the unauthorized \gls*{uav}, various interdiction methods can be deployed, such as jamming the remote control or GPS links of the target \gls*{uav} or throwing a net to render the target \gls*{uav} inoperable.
Such complex scenarios can be generated in \gls*{dt} for training and validation of \gls*{ai}-aided decision mechanisms.

\subsection{Federated Learning and Over-the-air Computation}

\Gls*{fl} is a well-studied distributed learning framework where model training is {\em federated} across multiple edge devices, sharing local information such as model parameters with an edge server for aggregation. 
In the context of \glspl*{uav}, FL is a promising approach to reduce \gls*{uav} energy consumption as the training can be handled with fewer communication instances between the sensors and \glspl*{uav}, as opposed to acquiring a larger number of data samples from many sensors, while also \emph{jointly} optimizing the locations of the UAVs~\cite{Chellapandi_2024}. 
To improve the spectrum utilization further for FL, a growing research area is \gls*{oac}, which takes advantage of the signal superposition property of wireless channels in the computation, e.g., average of local gradients~\cite{sahin_OACsurvey2023}. 
The remarkable gain obtained with \gls*{oac} is that the resource usage can be reduced to a one-time cost as all sensors transmit simultaneously with OAC.
Integrating such techniques into \glspl*{dt} can provide unparalleled insights and optimization opportunities, enabling \gls*{rw} deployment of FL in \glspl*{avn} with unprecedented efficiency and scalability. 
\subsection{Beamforming and Tracking with UAVs}

Millimeter wave (mmWave) communication systems require deploying large antenna arrays to achieve sufficient beamforming gains. 
Realizing these gains, however, requires mmWave transceivers to align their beams, typically through a beam-sweeping process over a pre-defined codebook.
These codebook sizes scale with the number of antennas resulting in high beam sweeping overhead that makes it hard for these systems to support highly mobile vehicular applications.  
In this context, \glspl*{dt} offer an interesting framework for assisting beam prediction and tracking in \gls*{uav} communication systems. 
For example, \glspl*{dt} can be utilized to generate high-fidelity site-specific datasets that capture the spatial and temporal characteristics of the target environment and \gls*{uav} mobility patterns~\cite{real-time-digital-twin}. 
These datasets can be used to develop and train the machine learning (ML) models, relaxing the need for collecting \gls*{rw} datasets. 
Further, in the operation phase, \glspl*{dt}--deployed at the edge or cloud--can be leveraged to provide near real-time situation awareness to the UAVs. 
This awareness may enable the UAVs to make proactive predictions about, for example, future blockages, and hence guide proactive beam switching decisions and avoid potential link disconnections. 

\subsection{Coverage Extension with UAVs}

With the always-increasing demand for connectivity, \glspl*{uav} can be used in conjunction with cellular towers and other \glspl*{uav} to increase the range of operation, enhance network capacity, and provide a more reliable and secure communication infrastructure.
However, manually controlling drones to extend coverage is impractical due to additional operational costs and compromised accuracy.
\Gls*{ai}-based solutions can be employed to optimize flight trajectories for ideal operation by designing and dynamically adapting them.
By leveraging \glspl*{dt} for training and validation of the \gls*{ai} models in this context, researchers can ensure accurate and safe testing of complex scenarios.


\section{AERPAW DT for Supporting AVN Research}
\gls*{aerpaw}~\cite{aerpaw_website} is a research platform funded by the National Science Foundation (NSF) to support research and development for \glspl*{avn}.
\gls*{aerpaw} offers a unique experimental setup combining \gls*{rw} outdoor wireless testing with a dedicated virtual development environment. 
By providing a tightly inter-connected virtual and physical platform for validating both the \gls*{p2v} and \gls*{v2p} connections, researchers can advance the field of AI-aided \glspl*{avn} and address the complexities associated with wireless networks and autonomous vehicle systems. 
Experiment workflow in the AERPAW platform, also highlighted in Fig.~\ref{fig:Testbed}, involves the development and initial testing in the \gls*{dt}, sandbox testing with Keysight Propsim channel emulator, and the final testing in the \gls*{pt}, where simulations can be used to improve the training and calibration process.

As a first step for the incoming experimenter, the \gls*{dt} of the \gls*{aerpaw} testbed is generated for the initial development based on the configuration provided by the experimenter.
This configuration includes the details about the experiment such as the numbers and types of portable nodes (e.g., \gls*{uav}, \gls*{ugv}), and fixed nodes (e.g., BSs).
\gls*{aerpaw} platform brings up a set of Docker containers for the experiment and initiates the \gls*{dt} components such as the channel and vehicle emulator.
Software for a typical experiment consists of three main components: 1) radio, 2) vehicle, and 3) application (e.g., network traffic generation) software.

\gls*{aerpaw}'s \gls*{dt} is a distinctive development environment that allows users to use the same software components as in the \gls*{pt}.
This is accomplished by using \gls*{sitl} vehicle emulator, virtual radio hardware stack, and the I/Q channel emulator developed by the \gls*{aerpaw} team.
All the emulation components are containerized and tightly connected to each other through \gls*{aerpaw}'s networking frame, enabling the experimenter to develop and test the software in the \gls*{dt} environment.
As a result, experiments can be transferred between different deployment options without making any adjustments to the developed software.
This feature allows experimenters to develop and test their software without the need for abstraction of the system and the hassle of the integration of the software into the real hardware.
Additionally, this allows the experiments to be deployed back to the \gls*{dt} for further testing and validation after the initial testing in the \gls*{pt}.
The experiment deployment cycle can be repeated multiple times to improve the software, \gls*{dt} and \gls*{ai} model based on the data collected from the \gls*{rw}.

\section{Case Study: AI-Aided Signal Source Localization with \glspl*{avn}}

In this section, we provide a case study to illustrate how a \gls*{dt} can be used to develop and test AI algorithms in \glspl*{avn}. 
\gls*{aerpaw} platform recently hosted the \gls*{aerpaw} Find-a-Rover (AFAR) Student Challenge~\cite{aerpaw_website}. 
In this competition, teams were tasked with localizing an \gls*{ugv}, similar to the use case discussed in Section~\ref{Sec:search_rescue}, using signals broadcasted at a narrowband (125 KHz) waveform by the UGV.
Due to the narrow bandwidth of the signal, the complex 3D propagation conditions, and the variations in the orientation and the tilt of the \gls*{uav}, the received signal strength is subject to deep fading effects that can be on the order of up to 60 dB as the \gls*{uav} navigates through the field. 
Participants developed their \gls*{ai}-based trajectory control and localization solutions based on signal strength variation in \gls*{aerpaw}'s \gls*{dt}. 
Upon completion of development, experiment containers from each team were deployed by AERPAW personnel in the \gls*{pt} for performance evaluation in the \gls*{rw} environment. 
As a representative example, this section provides the solution approach and representative results from one of the teams.   




\subsection{Development in AERPAW Digital Twin} 

\begin{figure*}[t]
    \centering
    \includegraphics[width=0.95\linewidth]{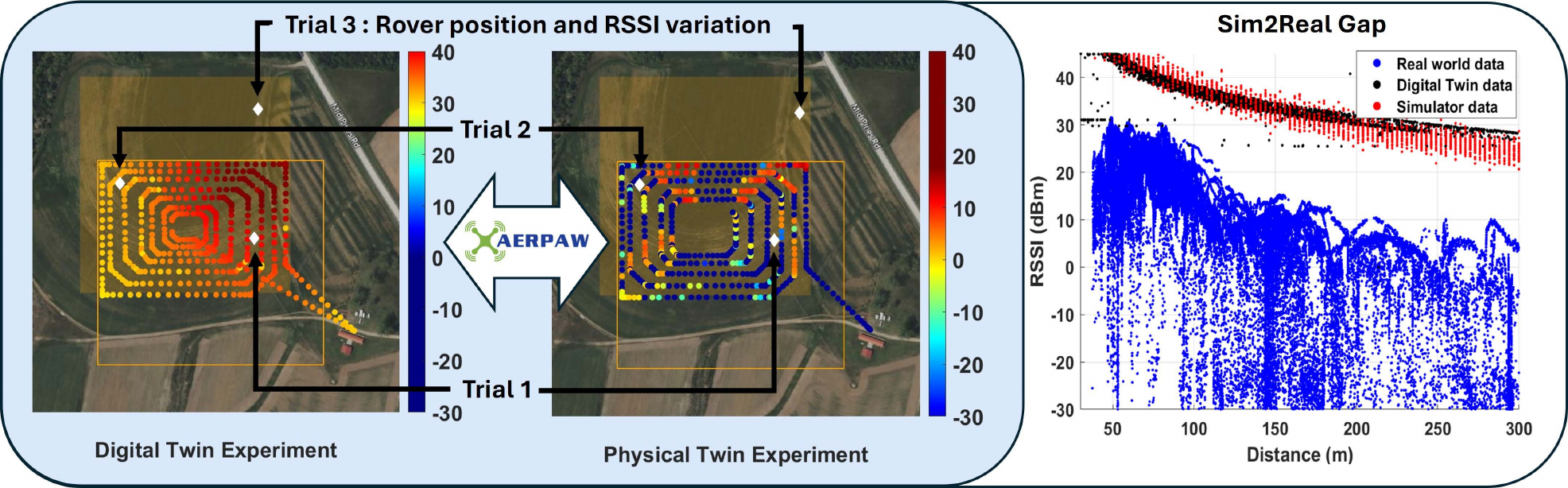}
    \caption{\gls*{rssi} variation in the \gls*{dt} and the \gls*{pt} across the UAV's trajectory. The deep fades in the \gls*{rw} environment adversely affect the localization performance if the training is performed exclusively based on the \gls*{dt} environment.}
    \label{fig:RSRP_data}\vspace{-4mm}
\end{figure*}

We first created a simulator of a representative scenario to generate a large-scale signal strength dataset, which is later used for training an \gls*{ai} algorithm. 
The channel propagation model in the simulator is derived from the two-ray path loss model with due consideration of antenna radiation patterns~\cite{afar_simulator}. 
Fig.~\ref{fig:RSRP_data} captures the fading nature of \gls*{rssi} values against the distance of separation between the \gls*{uav} and the \gls*{ugv}. 
The simulated model includes nodes that closely mimic the \gls*{dt} in the loss pattern, whereas the 
deep fades in the \gls*{rw} are considerably higher than those in the \gls*{aerpaw} \gls*{dt} environment.
This type of difference between \gls*{rw} testbeds and the models significantly affects the performance of systems developed based on statistical simulations. 
In this section, we first present the \gls*{rw} localization performance from systems developed in a simulator+\gls*{dt} environment and then augment the development cycle with \gls*{rw} data to demonstrate enhanced performance.  

\textit{\underline{Particle Filters}:} We explored two approaches for localizing the rover in the AFAR challenge. 
The first is a traditional technique based on particle filters (PFs).
PFs are well-suited for non-linear measurement models~\cite{particle_filter}, making them an attractive choice for the \gls*{rf} source localization task.
The PF is implemented as a recursive process with three steps: state update, measurement update, and particle redistribution.
The PF begins with a uniform distribution of particles posing as \gls*{ugv} positions across the search area. 
Next, the \gls*{uav} is allowed to traverse on a fixed trajectory to gather \gls*{rssi} measurements across the search area. 
At each \gls*{uav} position on the trajectory, sample \gls*{rssi} values are derived from all the particles using the path loss model present in the simulator. 
The sample values are compared against the current \gls*{rssi} measurement to derive their likelihood assuming an underlying Gaussian distribution for the deviation. 
At each iteration, the location estimate is obtained as a weighted average of all the particle positions with the weights defined by their respective normalized likelihood probabilities. 
These weights are also used to redistribute particles before the next iteration to avoid particle impoverishment.   

\textit{\underline{Convolutional Neural Networks}:} Deep Learning has proven its prowess in various domains, and its potential application in localization is a promising research avenue~\cite{kim2018scalable}. 
We introduce Finger-CNN as our second localization approach, which combines the strength of the Convolutional Neural Network (CNN) with the fingerprinting approach. 
This CNN-based algorithm is designed for classification tasks, facilitating the partitioning of the area of interest (AoI) into smaller regions and subsequently determining the presence or absence of the \gls*{rf} source within those defined regions. 
The CNN needs a 2D array as input, which is achieved by combining RSSI data with spatial coordinates of \gls*{uav} positions to form an image. 
A visual representation, shown in Fig.~\ref{fig:RSRP_data}, presents \gls*{uav} positions as points in a scatter plot, where each point is color-coded according to its respective RSSI value.
The AoI for the UGV is a rectangle divided into $C$ equal smaller rectangles. 
During training data generation, UGV is randomly positioned within AoI, and class labels are assigned based on which rectangle it falls into. 
During testing, the Finger-CNN predicts UGV's class, which is converted into longitude and latitude, providing precise location prediction within the AoI.

The Finger-CNN model consists of three stages: feature extraction, flattening, and classification. 
Feature extraction involves three convolutional layers with increasing complexity, each followed by max pooling. 
In the classification stage, the final layer consists of $C$ neurons with softmax activation, producing a probability distribution across the classes. The Finger-CNN model is implemented using JAX and Flax, which are optimized for high-performance numerical computing and large-scale machine learning. The total dataset was divided into three parts: $70\%$ for training, and $15\%$ each for testing and validation. The model was trained for 200 epochs with a learning rate of 1e-3, using an exponential decay learning rate scheduler that reduces the learning rate by a factor of 0.1, and a weight decay of 0.001 was applied.

\begin{figure}[t]
    \centering
    \includegraphics[width=\linewidth]{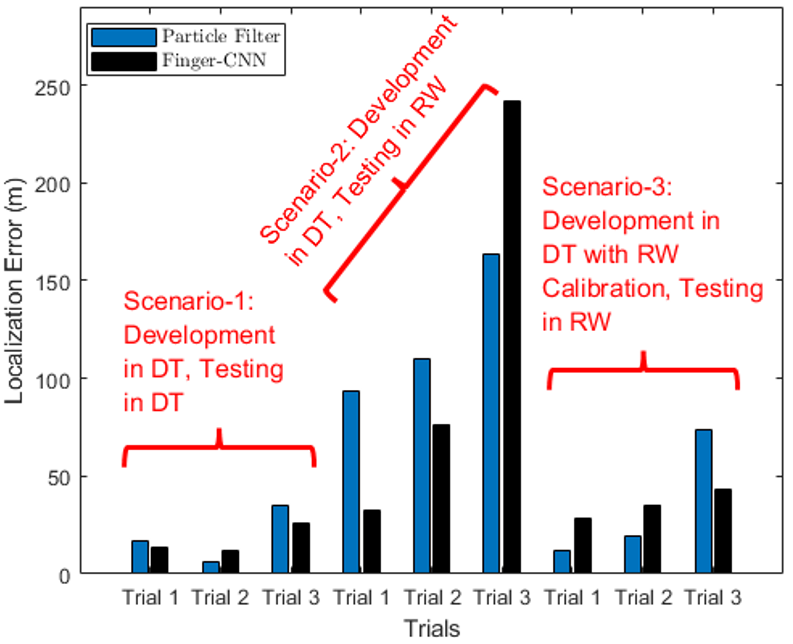}
        \caption{Localization results using particle filter and finger-CNN techniques. For improved performance, the development in the DT is also supported with \gls*{rw} data.}
    \label{fig:FPRresults}\vspace{-3mm}
\end{figure}

\subsection{Results and Discussion}

Fig.~\ref{fig:FPRresults} shows localization errors across 3 trials with varying rover positions. 
Trial~1 and Trial~2 are when the UGV is inside the UAV flight area, while Trial~3 represents the case when the UGV is outside this area (see Fig.~\ref{fig:RSRP_data}).
We present localization results for PF and Finger-CNN for three different scenarios: 1) development in a DT (Sim+DT) environment, and then tested in DT; 2) development in a DT (Sim+DT) environment, and then tested in \gls*{rw}; 3) development in a DT (Sim+DT) environment with calibration based on \gls*{rw} data, and then tested in \gls*{rw}.  
For Scenario-1 and Scenario-2, PF operates with a path loss model matching the DT and the Finger-CNN is trained with a simulator as suggested in Fig.~\ref{fig:Testbed} while fine-tuning of network parameters is done with the help of DT. When tested in the \gls*{rw} (Scenario-2), the Finger-CNN provides high accuracy in Trial~1 and Trial~2 owing to in-depth RSSI fingerprint knowledge of the UAV search area. 
On the other hand, Finger-CNN suffers from insufficient training data for the out-of-search-area case as seen with Trial~3. 
PF optimally tuned for noise removal is range agnostic and better estimates the rover position for Scenario-2. 
Overall, the localization errors are very high for Scenario-2 due to the unaccounted \gls*{rw} channel effects encountered for the first time when testing in the \gls*{pt}.

To improve the localization results we gather \gls*{rw} logs from all AFAR submission teams and use them to finetune our simulator without overfitting our models. 
We redevelop the methods as before with changes only to the training data. The modified training data now accounted for a wider spread in RSSI values, deep fades, and a steeper path loss model. The redeveloped estimation methods are tested with the data logs for each trial. 
The new results, shown as Scenario-3 in Fig.~\ref{fig:FPRresults} have lower error margins as the PF is better tuned for the offset in RSSI values, and the Finger-CNN can generalize better even for the out-of-search area scenario in the third trial. 
While we do not consider real-time optimization of the UAV trajectory in this work, this was used by several AFAR teams~\cite{aerpaw_website} and further gains in localization accuracy are possible. 

\section{Conclusion}

In this article, we explore the role of \glspl*{dt} as a development environment 
for \gls*{ai}-aided \glspl*{avn}.
We study the potential of using \glspl*{dt} to improve system functionality in various use cases that involve optimization of AV trajectories and wireless parameters. Moreover, we discuss the development environments commonly used for \glspl*{avn}, positioning \gls*{dt} as a crucial component within this pipeline.
At last, we provide an in-depth case study using \gls*{aerpaw}'s \gls*{dt} and \gls*{pt}, as well as the development cycle, for signal source localization employing \gls*{ai} techniques. 
Our results demonstrate that a \gls*{dt} development environment that uses \gls{rw} data for calibration, supported by additional simulations for fine-tuning parameters, can serve as an ideal platform for developing and testing AI-aided AVN technologies.
Future research directions include the development of more complex AI algorithms for \glspl*{avn} within \gls*{dt}, such as federated learning and over-the-air computation, and the integration of these algorithms into the \gls*{aerpaw} platform for further validation and testing.


\vspace*{-1mm}
\section*{Biographies}
\vspace*{-12mm}
\begin{IEEEbiographynophoto}{An\i l G\"urses}
    is a PhD student in Electrical Engineering at NC State University.
    His research focuses on digital twins and wireless channel emulation.
\end{IEEEbiographynophoto}
\vspace*{-12mm}
\begin{IEEEbiographynophoto}{Gautham Reddy}
    is a PhD student in Electrical Engineering at NC State University. His research focuses on 5G, network architectures, and ORAN.
\end{IEEEbiographynophoto}
\vspace*{-12mm}
\begin{IEEEbiographynophoto}{Saad Masrur}
    is a PhD student in Electrical Engineering at NC State University. His research focuses on wireless communication and ML.
\end{IEEEbiographynophoto}
\vspace*{-12mm}
\begin{IEEEbiographynophoto}{\"Ozg\"ur \"Ozdemir}
    is an Associate Research Professor at the Department of Electrical and Computer Engineering at NC State University. His research interests include mmWave channel sounding, SDRs, and UAV communications.
\end{IEEEbiographynophoto}
\vspace*{-12mm}
\begin{IEEEbiographynophoto}{\.{I}smail G\"uven\c{c}}
    is a Professor at the Department of Electrical and Computer Engineering at NC State University. 
    His recent research interests include 5G/6G wireless networks and UAV communications.
\end{IEEEbiographynophoto}
\vspace*{-12mm}
\begin{IEEEbiographynophoto}{Mihail L. Sichitiu}
    is a Professor at the Department of Electrical and Computer Engineering at NC State University. 
    His recent research interests include UAV networks and digital twins.
\end{IEEEbiographynophoto}
\vspace*{-12mm}
\begin{IEEEbiographynophoto}{Alphan \c{S}ahin}
    is an Assistant Professor at the University of South Carolina. His research is on signal processing techniques for the physical layer of wireless communication systems. 
\end{IEEEbiographynophoto}
\vspace*{-12mm}
\begin{IEEEbiographynophoto}{Ahmed Alkhateeb}
    is an Assistant Professor at ASU. He holds a PhD from UT Austin and has received awards including the 2016 IEEE Signal Processing Society Young Author Best Paper Award.
\end{IEEEbiographynophoto}
\vspace*{-12mm}
\begin{IEEEbiographynophoto}{Magreth Mushi}
    is AERPAW Platform Director. She contributed significantly to the development of AERPAW's network and the Digital Twin architecture. She holds a PhD from North Carolina State University. 
\end{IEEEbiographynophoto}
\vspace*{-12mm}
\begin{IEEEbiographynophoto}{Rudra Dutta}
    is a Professor at the Department of Electrical and Computer Engineering at NC State University. 
    His recent research interests include network orchestration and software-defined networks. 
\end{IEEEbiographynophoto}
\vspace*{-4mm}

\bibliographystyle{IEEEtran}
\bibliography{bibtex/bib/IEEEabrv,bibtex/bib/refs}

\begin{thebibliography}{10}
\providecommand{\url}[1]{#1}
\csname url@samestyle\endcsname
\providecommand{\newblock}{\relax}
\providecommand{\bibinfo}[2]{#2}
\providecommand{\BIBentrySTDinterwordspacing}{\spaceskip=0pt\relax}
\providecommand{\BIBentryALTinterwordstretchfactor}{4}
\providecommand{\BIBentryALTinterwordspacing}{\spaceskip=\fontdimen2\font plus
\BIBentryALTinterwordstretchfactor\fontdimen3\font minus \fontdimen4\font\relax}
\providecommand{\BIBforeignlanguage}[2]{{%
\expandafter\ifx\csname l@#1\endcsname\relax
\typeout{** WARNING: IEEEtran.bst: No hyphenation pattern has been}%
\typeout{** loaded for the language `#1'. Using the pattern for}%
\typeout{** the default language instead.}%
\else
\language=\csname l@#1\endcsname
\fi
#2}}
\providecommand{\BIBdecl}{\relax}
\BIBdecl

\bibitem{aerpaw_website}
\BIBentryALTinterwordspacing
{AERPAW}, ``{Aerial Experimentation and Research Platform for Advanced Wireless},'' 2024. [Online]. Available: \url{https://aerpaw.org/}
\BIBentrySTDinterwordspacing

\bibitem{dtn_offloading}
Y.~Dai, K.~Zhang, S.~Maharjan, and Y.~Zhang, ``Deep reinforcement learning for stochastic computation offloading in digital twin networks,'' \emph{IEEE Trans. Industrial Informatics}, vol.~17, no.~7, pp. 4968--4977, 2021.

\bibitem{dtn_6g_vision}
L.~U. Khan, W.~Saad, D.~Niyato, Z.~Han, and C.~S. Hong, ``Digital-twin-enabled {6G}: Vision, architectural trends, and future directions,'' \emph{IEEE Communications Magazine}, vol.~60, no.~1, pp. 74--80, 2022.

\bibitem{itu-tDTN}
ITU-T, ``{Digital twin network – Requirements and architecture},'' 2 2022, recommendation ITU-T Y.3090.

\bibitem{jones2020characterising}
D.~Jones, C.~Snider, A.~Nassehi, J.~Yon, and B.~Hicks, ``Characterising the digital twin: A systematic literature review,'' \emph{CIRP Journal of Manufacturing Science and Technology}, vol.~29, May 2020.

\bibitem{dtn_challenges}
A.~Masaracchia, V.~Sharma, B.~Canberk, O.~A. Dobre, and T.~Q. Duong, ``Digital twin for 6g: Taxonomy, research challenges, and the road ahead,'' \emph{IEEE Open J. Commun. Society}, pp. 2137--2150, 2022.

\bibitem{minimize_data_collection}
Y.~Wang, Z.~Hu, X.~Wen, Z.~Lu, and J.~Miao, ``Minimizing data collection time with collaborative {UAVs} in wireless sensor networks,'' \emph{IEEE Access}, vol.~8, pp. 98\,659--98\,669, 2020.

\bibitem{uav_assisted_networks}
X.~Liu, Y.~Liu, Y.~Chen, and L.~Hanzo, ``Trajectory design and power control for multi-{UAV} assisted wireless networks: A machine learning approach,'' \emph{IEEE Trans. Veh. Technol.}, vol.~68, no.~8, 2019.

\bibitem{koohifar2018autonomous}
F.~Koohifar, I.~Guvenc, and M.~L. Sichitiu, ``Autonomous tracking of intermittent {RF} source using a {UAV} swarm,'' \emph{IEEE Access}, vol.~6, pp. 15\,884--15\,897, 2018.

\bibitem{Chellapandi_2024}
V.~P. Chellapandi, L.~Yuan, C.~G. Brinton, S.~H. Żak, and Z.~Wang, ``Federated learning for connected and automated vehicles: A survey of existing approaches and challenges,'' \emph{IEEE Trans. Intelligent Veh.}, vol.~9, no.~1, pp. 119--137, 2024.

\bibitem{sahin_OACsurvey2023}
A.~\c{S}ahin and R.~Yang, ``A survey on over-the-air computation,'' \emph{IEEE Commun. Surveys \& Tutorials}, vol.~25, no.~3, pp. 1877--1908, 2023.

\bibitem{real-time-digital-twin}
A.~Alkhateeb, S.~Jiang, and G.~Charan, ``Real-time digital twins: Vision and research directions for 6{G} and beyond,'' \emph{IEEE Commun. Mag.}, vol.~61, no.~11, pp. 128--134, 2023.

\bibitem{afar_simulator}
S.~J. Maeng, H.~Kwon, O.~Ozdemir, and I.~Guvenc, ``Impact of {3-D} antenna radiation pattern in {UAV} air-to-ground path loss modeling and {RSRP}-based localization in rural area,'' \emph{IEEE Open J. Antennas Propag.}, vol.~4, pp. 1029--1043, 2023.

\bibitem{particle_filter}
F.~Gustafsson, F.~Gunnarsson, N.~Bergman, U.~Forssell, J.~Jansson, R.~Karlsson, and P.-J. Nordlund, ``Particle filters for positioning, navigation, and tracking,'' \emph{IEEE Trans. Sig. Proc.}, pp. 425--437, 2002.

\bibitem{kim2018scalable}
K.~S. Kim, S.~Lee, and K.~Huang, ``A scalable deep neural network architecture for multi-building and multi-floor indoor localization based on {Wi-Fi} fingerprinting,'' \emph{Big Data Analytics}, vol.~3, pp. 1--17, 2018.

\end{thebibliography}

\end{document}